\documentclass{PoS}
\newcommand{\beq}{\begin{equation}}
\newcommand{\eeq}{\end{equation}}
\newcommand{\beqn}{\begin{eqnarray}}
\newcommand{\eeqn}{\end{eqnarray}}
\newcommand{\nn}{\nonumber}

\title{Sea quark QED effects and twisted mass fermions}

\ShortTitle{Sea quark QED effects and twisted mass fermions}

\author{\speaker{R.\ Frezzotti} \\
       Dipartimento di Fisica, Universit\`a di  Roma ``{\it Tor Vergata}'' and INFN, Sezione di Roma 2\\
        E-mail: \email{frezzotti@roma2.infn.it}}
\author{{G.C.\ Rossi} \\ 
       Dipartimento di Fisica, Universit\`a di  Roma ``{\it Tor Vergata}'' and INFN, Sezione di Roma 2\\
       Centro Fermi - Museo Storico della Fisica, Piazza del Viminale 1 - 00184 Roma, Italy\\
        E-mail: \email{rossig@roma2.infn.it}}

\author{{N.\ Tantalo}\\ 
       Dipartimento di Fisica, Universit\`a di  Roma ``{\it Tor Vergata}'' and INFN, Sezione di Roma 2\\
        E-mail: \email{nazario.tantalo@roma2.infn.it}}

\abstract{
We show that maximally twisted mass fermions can be employed to regularize on the lattice
the fully unquenched QCD+QED theory with vanishing $\theta$-term. We discuss how the critical
mass of the up and down quarks can be conveniently determined beyond the electroquenched
approximation by imposing that certain symmetries of continuum QCD+QED, which are broken by
Wilson terms, get restored in the continuum limit. A mixed action setup is outlined that 
allows to extend beyond the electroquenched approximation the computation (with only O($a^2$) artifacts)
of the leading isospin breaking corrections to physical observables using the RM123 method 
and (pure QCD) ETMC gauge ensembles with $N_f=2+1+1$ dynamical quark flavours.
}

\FullConference{34th annual International Symposium on Lattice Field Theory\\
		24-30 July 2016\\
		University of Southampton, UK}

\begin{document}

\section{Introduction}
\label{sec:INTRO}

In refs.~\cite{deDivitiis:2011eh,deDivitiis:2013xla} a strategy for evaluating the
leading isospin breaking (LIB) corrections to hadronic quantities has been proposed. 
It is based on expanding the full Q(C+E)D observables to first order in powers of the small quantities $(m_d-m_u)/\Lambda_{QCD}$ and $\alpha_{em}$ (so-called RM123 approach). In this way the computational task is reduced to that of evaluating hadronic correlators with insertions of electromagnetic (e.m.) currents or quark scalar densities in the theory with no electromagnetism and no $u$-$d$ mass splitting (isosymmetric theory) .

In the same papers the viability of the RM123 approach was tested by evaluating the LIB corrections to hadron masses as well as the Dashen's theorem breaking parameter $\epsilon_\gamma$ using the $N_f=2$ ensembles generated by ETMC~\cite{Blossier:2010cr,Baron:2009wt}. 
The study of LIB effects in the leptonic meson decay rates, which was started
in ref.~\cite{deDivitiis:2013xla} as far as the correction $\propto (m_d-m_u)$ is concerned,
has been recently extended to the LIB e.m.\ corrections (see talks by Tantalo~\cite{NT1} 
and Simula~\cite{Lubicz:2016mpj}) by following
the general strategy of refs.~\cite{Carrasco:2015xwa,Lubicz:2016xro}, which allows to keep under control 
the infrared divergencies arising in the intermediate steps of the calculation.

All these investigations have been carried out so far in the electro-quenched approximation,
 i.e.\
by treating sea quarks as if they were electrically neutral (diagrammatically this means 
neglecting all contributions with photons attached to sea quark loops). Though this is
a reasonable first approximation, it appears difficult to reliably control the systematic
error it induces on the computed LIB effects. 
Here we discuss how twisted mass LQCD~\cite{Frezzotti:2003ni,Frezzotti:2004wz,Frezzotti:2003xj} can be 
conveniently combined with the RM123 approach in order to evaluate~\footnote{From the 
numerical viewpoint it is assumed (with no discussion here) that suitable methods are
employed to evaluate the fermionically disconnected diagrams that arise
when accounting for electro-unquenched LIB effects.} LIB corrections to
hadronic observables {\em beyond the electro-quenched approximation} in the theory 
with dynamical $u$, $d$ as well as $s$ and $c$ quarks.

\section{A lattice regularization of Q(C+E)D with maximally twisted Wilson fermions}
\label{sec:ANOMALIES}

The use of maximally twisted Wilson fermions allows to avoid O($a$) lattice
artifacts in physical observables at the price of introducing parity and isospin 
breakings at finite lattice spacing, which come on top of the physical isospin 
violations. One may thus worry that in the presence of e.m. interactions 
a delicate tuning of bare mass parameters is needed in order to obtain a 
continuum effective action
with no strong and e.m.\ $\theta$-terms.
We show here that this is not the case if one works at maximal twist, i.e.\ with
$M_0 = M_{cr}$ for each quark pair. 


We discuss the conceptual point in LQ(C+E)D with $u$ and $d$ (non-degenerate) quarks. 
The quark lattice action of Q(C+E)D for a maximally twisted isospin doublet
$\psi=(\psi_u,\psi_d)$ reads~\cite{Horkel:2015xha}
{\small{
\beqn
S_{\rm{F}}^{\rm{Q(C+E)D}}(\psi, \bar\psi, U,E)=a^4 \sum_x\,\bar\psi(x)\Big{[}
\gamma\cdot{\widetilde\nabla}-i\gamma_5{\tau_3 W_{\rm{cr}} }
+\mu+{\tau_3} \epsilon\Big{]}\psi(x) \label{LATACT}\eeqn
\vspace{-.7cm}
\beqn
\hspace{-.8cm}&&\gamma\cdot {\widetilde\nabla} \equiv
\frac{1}{2}\sum_\mu\gamma_\mu(\nabla^\star_\mu+\nabla_\mu)\, ,
\qquad \quad {W_{\rm{cr}} }
\equiv-\frac{a}{2}\sum_\mu{\nabla^\star_\mu\nabla_\mu} +{M_{\rm{cr}} }
\nn\\
\hspace{-.8cm}&&{\nabla_\mu} \psi(x) \equiv \frac{1}{a}
\Big{[}\ {{\cal U}_\mu }(x)\psi(x+a\hat\mu) - \psi(x) \Big{]} \, , \quad
{ \nabla_\mu^* } \psi(x) \equiv \frac{1}{a}
\Big{[} \psi(x) - { {\cal U}^\dagger_\mu }(x-a\hat\mu) \psi(x-a\hat\mu) \Big{]} \nn
\eeqn
\vspace{-.7cm}
\beqn
\hspace{-1.2cm}&& {{\cal U}_\mu }(x) =
{E_\mu }(x) U_\mu(x) \, , \qquad \quad
{ E_\mu} (x) = \frac{1\!\!1+{\tau_3}}{2} e^{ie {q_{u}}  {{\cal A}_\mu} (x)} + \frac{1\!\!1-{\tau_3}}{2}  e^{ie {q_{d}}  {{\cal A}_\mu} (x)}\, , \label{TLQCED}
\eeqn}}
where $U_\mu(x)$ and ${ E_\mu} (x)$ are the strong and e.m.\ (non-compact QED) gauge links. The
flavour structure of the latter is due to the unequal electric charges of {\it up}
and {\it down} quarks. In flavour space the charge operator reads
$Q=e(1\!\!\!1/6 + \tau_3/2)$, where $e$ is the electric charge. 
Consequently also the critical mass counterterm, $M_{cr}$, will take the diagonal matrix form
{\small 
\beqn
&& \hspace{-1.5cm} M_{cr} = {M_{cr}^u}
\frac{{ 1\!\!\!1} +{ \tau_3}}{2}
  + { M_{cr}^d} \frac{{1\!\!1}-{ \tau_3}}{2} \equiv
  m_{cr}{ 1\!\!\!1} +\tilde m_{cr}{ \tau_3 } \label{MCRF}\, .
\eeqn 
}
We stress that in eq.~(\ref{LATACT}) the critical Wilson term,
$ - \bar\psi i \gamma_5 \tau^3 W_{cr} \psi$, is chirally twisted in the $\tau_3$
``isospin direction'' to comply with e.m.\ gauge invariance, which leads to
a complex quark determinant (see sect.~\ref{sec:SISTRA} on how the resulting
problem can be circumvented in the RM123 approach).

For convenience we separate out quark flavours and rewrite eq.~(\ref{LATACT}) in the case of generic twist angles,
called $\theta_u$ and $\theta_d$ (maximal twist is recovered e.g.\ for 
$\theta_u=-\theta_d=\pi/2$), obtaining
{\small{\beqn
\hspace{-.8cm}&&S^{{LQ(C+E)D}}(\psi, \bar\psi, U,E)=
a^4 \sum_x\,\Big[ \, FF|_E(x) + {\rm tr}(GG)|_U(x) \,\Big] +
\nn\\
\hspace{-.8cm}&&+a^4 \sum_x\,\bar\psi_u(x)\Big{[}\gamma\cdot\widetilde\nabla^u+e^{-i\theta_u\gamma_5}W_{\rm{cr}}^u+M_u\Big{]}\psi_u(x)+
a^4\, \sum_x\,\bar\psi_d(x)\Big{[}\gamma\cdot\widetilde\nabla^d+e^{-i\theta_d\gamma_5}W_{\rm{cr}}^d+M_d\Big{]}\psi_d(x)\, ,\label{UD}\eeqn}}
\noindent
where $M_{u/d} = \mu \pm \epsilon$ are bare quark masses, 
$W_{\rm cr}^f = - \frac{a}{2} \sum_\mu \nabla^{*f} \nabla^f + M_{cr}^f$ and
a superscript $^u$ or $^d$ is attached to the covariant derivatives 
because $q_{u}\neq q_{d}$.  
Eq.~(\ref{UD}) shows that $S^{{LQ(C+E)D}}$ is a periodic function of $\theta_u$ 
and $\theta_d$. Since {\em lattice} singlet axial rotations are not anomalous, 
twist phases can be freely moved from the Wilson to the mass terms by means of 
the axial rotations 
\vspace{-.1cm}
{\small{\beqn
&&{\mbox{U}}_A^u(1):\quad \psi_u =e^{i\gamma_5\theta_u/2}\chi_u \, \qquad \bar \psi_u =\bar \chi_u e^{i\gamma_5\theta_u/2} \label{ROTU}\\
&&{\mbox{U}}_A^d(1):\quad\psi_d =e^{i\gamma_5\theta_d/2}\chi_d \, \qquad \bar \psi_d =\bar \chi_d e^{i\gamma_5\theta_d/2} \label{ROTD}
\eeqn}}
by which the lattice action~(\ref{UD}) can be brought into the form
(with untwisted Wilson terms $W_{\rm{cr}}^{u,d}$)
{\small{\beqn
&S^{{LQ(C+E)D}}(\chi, \bar\chi, U,E)=
a^4 \sum_x\,\Big[ \, FF|_E + {\rm tr}(GG)|_U + \!
\sum_{f=u,d} \! \Big( \bar\chi_f \gamma\cdot\widetilde\nabla^f \chi_f +
\bar\chi_f W_{\rm{cr}}^f \chi_f \Big) \Big] (x) +
{\mbox {MTs}} \; , 
\label{ROTATED} \\
& {\mbox {MTs}} = a^4 \sum_x\, \sum_{f=u,d} \bar\chi_f(x)
e^{i\theta_f\gamma_5}M_f \chi_f(x) \equiv a^4 \sum_x\, 
\Big[ m_u S_\chi^u(x) + i \mu_u P_\chi^u + m_d S_\chi^d(x) + i \mu_d P_\chi^d \Big]  
\, , \label{MTUD}
\eeqn}}
%
%
where we have introduced the lattice bare quark mass parameters and density operators 
\beq
m_f \equiv M_f \cos \theta_f \; , \quad \mu_f \equiv M_f \sin \theta_f \quad\; {\rm and} \quad\;
S_\chi^f \equiv\bar \chi^f \chi^f \; , \quad P_\chi^f \equiv\bar \chi^f \gamma_5\chi^f\; ,
\quad\; f = u,d \, . 
\label{DEFdens}
\eeq
Symmetries of the lattice action~(\ref{ROTATED}) 
and power counting arguments imply~\cite{PREP} that the local effective action
of the corresponding continuum theory can be written in the form
{\small{\beqn
&&S_{cont}^{{Q(C+E)D}}(\chi^c, \bar\chi^c, A,{\cal A})=
 \int dx\,\Big[ \, FF|_{\cal A}(x) + {\rm tr}(GG)|_{A}(x) 
 + \sum_{f=u,d}\, \bar\chi_f \gamma \cdot D^f|_{A, \cal A} \chi^f   \,\Big] 
+ {\rm MTs} \; ,  \label{CONT-ROTATED}\\
&& {\rm MTs} \, = \, \int dx\, \Big[ \, \hat m_u \hat S_\chi^u + 
i \hat \mu_u \hat P_\chi^u + \hat m_d \hat S_\chi^d + 
i \hat \mu_d \hat P_\chi^d  \, \Big]
\, ,
\eeqn}}
where $\chi_f$, $\bar \chi_f$, ${\cal A}$ (photon) and $A$ (gluon) 
stand now 
for suitably normalized continuum fields and
{\small{\beqn
&& \hat m_f = m_f^{\rm sub} (Z_S^f)^{-1} \; ,\quad \hat \mu_f = \mu_f^{\rm sub} (Z_P^f)^{-1}
\quad {\rm and} \quad
\hat S_\chi^f = Z_S^f S_\chi^{f, \rm sub} \; ,\quad \hat P_\chi^f =Z_P^f P_\chi^{f, \rm sub} 
\; , \quad f=u,d \quad
\label{MASSREN1} 
\eeqn}} 
denote the renormalized mass parameters and quark densities. 
%
Due to differencies in the renormalization constants of the relevant 
flavour singlet and non-singlet densities which start at 2-loop level 
and imply $ S_\chi^{u(d), \rm sub} = 
S_\chi^{u(d)} (1+ \rho^S_{u(d)})  + S_\chi^{d(u)} \tilde\rho^S_{d(u)}  $
as well as $ P_\chi^{u(d), \rm sub} = 
P_\chi^{u(d)} (1+ \rho^P_{u(d)})  + P_\chi^{d(u)} \tilde\rho^P_{d(u)}  $, 
the multiplicatively renormalizable mass parameters take
(see e.g.\ ref.~\cite{Horkel:2015xha}) the form
{\small{\beqn
&& m_{u(d)}^{\rm sub} = m_{u(d)} (1+ \rho^m_{u(d)}) + m_{d(u)} \tilde\rho^m_{d(u)}
\;, \quad
\mu_{u(d)}^{\rm sub}= \mu_{u(d)} (1+ \rho^\mu_{u(d)})+ \mu_{d(u)} \tilde\rho^\mu_{d(u)} 
\; . \label{MASSREN2}
\eeqn}}
%
%
Noting that the twisted mass terms $i (\hat\mu_{u} \hat P^u_\chi + 
\hat\mu_{d} \hat P^d_\chi)$ and the action terms ${\rm tr}(\tilde{G}G)$ and $\tilde F F$
(which can only appear with coefficients $\propto \hat \theta_{f}$ being odd
functions of $\hat\mu_{f}/\hat m_{f}$, $f=u,d$) are {\em not} independent operators in 
the {\em continuum} theory, as they are related by anomalous axial U(1)
rotations of the quark fields (see below), one cheks that the form~(\ref{CONT-ROTATED}) 
of the continuum action is indeed the most general one. 
%
Trading $\hat m_f$ and $\hat \mu_f$ for the alternative renormalized parameters
$\hat M_f$ and $\hat \theta_f$ ($f=u,d$),
{\small
\beqn
&\hat M_u = \sqrt{ \hat m_u^2 + \hat\mu_u^2} \; , \quad
\hat M_d = \sqrt{ \hat m_d^2 + \hat\mu_d^2} \; , \quad 
& \tan \hat\theta_u = \frac{Z_S^u}{Z_P^u}\,\tan \theta_u \; ,
\quad \tan \hat\theta_d = \frac{Z_S^d}{Z_P^d}\,\tan \theta_d\, , \quad\label{REWT} 
\eeqn
}
the continuum theory effective action~(\ref{CONT-ROTATED}) is rewritten as
{\small{\beqn
&&S_{cont}^{{Q(C+E)D}}(\chi^c, \bar\chi^c, A,{\cal A})=
 \int dx\,\Big[ \, FF|_{\cal A}(x) + {\rm tr}(GG)|_{A}(x) \,\Big] +\label{CONT}\\
&&+\int dx\,\bar\chi_u(x)\Big{[}\gamma\cdot D^u+e^{i\hat\theta_u\gamma_5}\hat M_u\Big{]}\chi_u(x)
+\int dx\,\bar\chi_d(x)\Big{[}\gamma\cdot D^d+e^{i\hat \theta_d\gamma_5}\hat M_d\Big{]}\chi_d(x)\nn  \, .
\eeqn}}
In the formal continuum theory the analogs of the U(1)-axial rotations~(\ref{ROTU}) 
and~(\ref{ROTD}) are anomalous, so the effective action~(\ref{CONT}) can be 
equivalently cast into the form
{\small{\beqn
&&S_{cont}^{{Q(C+E)D}}(\psi, \bar\psi, A,{\cal A})=
 \int dx\,\Big[ \, FF|_{\cal A}(x) + {\rm tr}(GG)|_{A}(x) \,\Big] +\label{CONT2}\\
&&+\int dx\,\bar\psi_u(x)\Big{[}\gamma\cdot D^u+\hat M_u\Big{]}\psi_u(x)
+\int dx\,\bar\psi_d(x)\Big{[}\gamma\cdot D^d+\hat M_d\Big{]}\psi_d(x)+\nn\\
&&+\frac{i}{32\pi^2}(\hat\theta_u+\hat\theta_d)\int dx \Big[ e^2 F\tilde F |_{\cal A}(x) +g^2 {\rm tr}(G\tilde G)|_{A}(x)  \Big] \, ,\nn
\eeqn}}
where
{\small 
\beq
\tilde F_{\mu\nu}=\frac{1}{2}\epsilon_{\mu\nu\rho\sigma}F_{\rho\sigma} \qquad \tilde G^a_{\mu\nu}=\frac{1}{2}\epsilon_{\mu\nu\rho\sigma}G^a_{\rho\sigma} \, .
\label{TLDE}
\eeq
}
Eq.~(\ref{CONT2}) shows that the lattice action~(\ref{UD}) leads to a 
continuum effective theory with vacuum-angle $\hat\theta_u+\hat\theta_d$.
In general, $\hat\theta_u+\hat\theta_d$ is not simply $\propto \theta_u+\theta_d$ 
due to e.m.\ interactions implying $\frac{Z_S^u}{Z_P^u} \neq \frac{Z_S^d}{Z_P^d}$.
However from eq.~(\ref{REWT}) one checks that at maximal twist (e.g. $\theta_u
= - \theta_d = \frac{\pi}{2}$) the continuum effective theory has $\hat\theta_u
= - \hat\theta_d = \frac{\pi}{2}$, hence vanishing vacuum-angle and no undesired
parity violation. 
In other words once $M_{cr}$ is determined we get Q(C+E)D for two non-degenerate quarks. 

\section{Fixing the critical mass in the twisted lattice Q(C+E)D theory}
\label{sec:TLQCED}
In our setting with two non-degenerate flavours we 
first have to non-perturbatively determine the critical mass,
$M_{cr}$ (see eq.\ref{MCRF}), appearing in eq.~(\ref{LATACT}).
As we shall work at first order in $\alpha_{em}$, we need to accordingly expand 
the parameters $m_{cr}$ and $\tilde m_{cr}$ of eq.(\ref{MCRF}) 
in powers of $\alpha_{em}$, obtaining 
{\small
\beqn
&& \hspace{-1.5cm} 
m_{cr}= m_{cr}^{LQCD}+ \alpha_{em} \frac{ \delta_{em}({\small g^2})}{a} +{\mbox O}(\alpha_{em}^2) 
\qquad \tilde m_{cr}=\alpha_{em} \frac{ \tilde \delta_{em}({\small g^2}) }{a} +{\mbox O}(\alpha_{em}^2) \, ,\label{MCR2}
\eeqn
}
where $\,m_{cr}^{LQCD}=w(g^2)/a $ is the critical mass of the isosymmetric theory.
From now on a suitable e.m.\ gauge fixing and procedure (e.g.\ QED$_L$) for
removal of the photon zero mode are assumed.  
Following the strategy outlined in~\cite{Bochicchio:1985xa,Frezzotti:2003ni,Frezzotti:2003xj} one can determine $M_{cr}$ by enforcing the chiral WTIs of Q(C+E)D.
For instance, with obvious notations for quark bilinears in $\psi$-basis
(e.g.\ $P^{1,2,3} = \bar\psi \gamma_5 \frac{\tau^{1,2,3}}{2} \psi$) and
$\delta q \equiv q_{u}-q_{d}=1$ a way to fix
%
%
$M_{cr}$ might be to enforce the formal continuum relations 
{\small{\beqn
\hspace{-.8cm}&&\langle \partial_\mu {V_\mu^1}(x) P^2(0)\rangle +e\,\delta q\, \langle  {{\cal A}_\mu} { V^2_\mu}(x)P^2(0)\rangle -2i \epsilon \langle {S^2}(x) P^2(0)\rangle =0\label{MCRV}\\
\hspace{-.8cm}&&\langle \partial_\mu {A_\mu^1}(x) S^1(0)\rangle -e\,\delta q\, \langle  {{\cal A}_\mu} { A^2_\mu}(x)S^1(0)\rangle -2\mu \langle { P^1} (x)S^1(0)\rangle =0\label{MCRA}\, .
\eeqn}}
This procedure is rather practical as the flavour structure of the two relations
is such that, once $m_{cr}^{LQCD}$ is known, it is possible to separately 
determine $\delta_{em}$ (from eq.~(\ref{MCRV})) and $\tilde\delta_{em}$ (from 
eq.~(\ref{MCRA}))~\footnote{Another way to fix $M_{cr}$ is to impose one chiral WTI (as was done in the electroquenched approximation in~\cite{deDivitiis:2013xla}) and then take as second condition the one arising from the minimization of $m_\pi^{\pm}$ with respect to $m_{u,d}$ (as is detailed in~\cite{Horkel:2015xha}).}.

A look at eqs.~(\ref{MCRV})) and~(\ref{MCRA}) shows that only parity violating correlators come into play. This fact together with a reconsideration of the condition usually employed in LQCD to determine the critical mass suggests a numerically simpler way to fix $m_{cr}$ and $\tilde m_{cr}$.

To explain the idea we first discuss the situation one meets in twisted mass LQCD. 
In this case the LQCD Symanzik effective action~\cite{SYM} for 
$m_0$ out of its critical value takes the form 
{\small{\beq \Gamma_{LQCD}= \int d^4 x  \,\{ L_4^{QCD}(x)
 +{ (w(g^2)/a-m_0)}\, [\bar\psi i\gamma_5\tau_3 \psi](x)
 + {a}L_5(x) +\ldots  \} \, .\label{GQCD}
 \eeq}}
\noindent
The undesired term $\propto [\bar\psi i\gamma_5\tau_3 \psi]$ in~(\ref{GQCD})
can be eliminated from $\Gamma_{LQCD}$ by enforcing the condition 
$\sum_{\vec{x}} \langle V_0^1(x) P^2(0) \rangle_{M_0}^{latt} =0$ 
(for $x_0 \gg a$)~\footnote{This condition in Wilson twisted mass LQCD returns what in ref.~\cite{Frezzotti:2005gi} 
was called the ``optimal'' critical mass.}. 
Indeed from the Symanzik expansion we obtain 
{\small{
\beqn
&&\langle V_0^1(x) P^2(0)\rangle^{latt}_{m_0}  = 
\langle V_0^1(x) P^2(0)\rangle^{L_4^{QCD}}  +
{ (w(g^2)/a - m_0)} \, \int \!d^4 y \langle V_0^1(x) P^2(0)   
[\bar\psi i\gamma_5\tau_3 \psi](y)\rangle^{L_4^{QCD}} + {\mbox{O}}({a})=\nn\\
&&={ (w(g^2)/a - m_0)} \, \int \!d^4 y \langle V_0^1(x) P^2(0)   
[\bar\psi i\gamma_5\tau_3 \psi](y)\rangle^{L_4^{QCD}} + {\mbox{O}}({a})\, ,
\label{QCDMCR}
\eeqn}}
where last equality follows from $\langle V_0^1(x) P^2(0)\rangle^{L_4^{QCD}}\!\!=0 $ 
in the parity-and-flavour invariant $L_4^{QCD}$.

To determine $M_{cr}$ in LQ(C+E)D we can use a similar method.
Working for simplicity with $\epsilon =0$ in~(\ref{LATACT})  
the Symanzik low energy action of LQ(C+E)D 
at $M_0 \equiv m_0 1\!\!1 + \tilde m_0 \tau^3 \neq M_{cr}$ reads
{\small{\beq \Gamma_{LQ(C+E)D}=\int d^4 x\{ L_{4}^{{Q(C+E)D}}(x)
 +(m_{cr}-m_0)\, [\bar\psi i\gamma_5\tau_3 \psi](y)
 +(\tilde m_{cr}- \tilde m_0)\, [\bar\psi i\gamma_5 \psi](x)
 +{a}L_5(x) +\ldots\}\, , \label{GQCED}
 \eeq}}
where we are simultaneously expanding in $m_0-m_{cr}$, $ \tilde m_0 - \tilde m_{cr}$ and
$a$ around the continuum theory $L_{4}^{{Q(C+E)D}}$ with $\hat m_u = \hat m_d =0$
(but $\hat \mu_u >0$, $\hat \mu_d <0$ and $\alpha_{em} \neq 0$).
Our claim is that $m_{cr}$ and $\tilde m_{cr}$ can be determined to first 
order in $\alpha_{em}$ by enforcing the conditions\\
\vspace*{-.3cm}
{\small 
\beqn 
&&\sum_{\vec x} \, \langle V^1_0(x)P^2(0)\rangle^{latt}_{M_0}=0 
\, , \qquad \sum_{\vec x} \, \langle S^1(x)P^1(0)\rangle^{latt}_{M_0}=0 
\, , \qquad x_0 \gg a \, . \label{CC1}
\eeqn
}
In fact, based on the effective action~(\ref{GQCED}) and using parity invariance
of $L_{4}^{Q(C+E)D} $, we can write
{\small{\beqn
\hspace{-.8cm}&&\langle V^1_0(x)P^2(0)\rangle^{latt}_{M_0} = {(m_{cr}^{LQCD}\!+ \!\alpha_{em} \delta_{em}a^{-1}\!-\!m_0)} \!\int\!\! d^4z \langle V^1_0(x)P^2(0) \bar\psi i\gamma_5\tau_3\psi(z)\rangle|^{L_{4}^{{Q(C+E)D}}} \!+\nn\\
\vspace*{-.6cm}
\hspace{-.8cm}&& \quad+ {(\alpha_{em} \tilde\delta_{em}a^{-1}\!-\!\tilde m_0)} { \!\int\!\! d^4z \langle V^1_0(x)P^2(0) \bar\psi i\gamma_5 \psi(z)\rangle|^{L_{4}^{{Q(C+E)D}}} } + \,{\mbox{O}}({a})
\label{VP}\\
\hspace{-.8cm}&&\langle S^1(x)P^1(0)\rangle^{latt}_{M_0} = {(\alpha_{em} \tilde\delta_{em}a^{-1}\!-\!\tilde m_0)} \!\int\!\! d^4z \langle S^1(x)P^1(0) \bar\psi i\gamma_5\psi(z)\rangle|^{L_{4}^{{Q(C+E)D}}} \!+\nn\\
\vspace*{-.6cm}
\hspace{-.8cm}&& \quad+ {(m_{cr}^{LQCD}\!+ \!\alpha_{em} \delta_{em}a^{-1}\!-\!m_0)} 
{ \!\int\!\! d^4z \langle S^1(x)P^1(0) \bar\psi i\gamma_5\tau_3\psi(z)\rangle|^{L_{4}^{{QC+E)D}}} } \!+ \,{\mbox{O}}({a}) \, .
\label{SP}
\eeqn}} 
\noindent
Assuming that $m_{cr}^{LQCD}$ is already known, for
our purposes  $\delta m_0 \equiv m_0-m_{cr}^{LQCD}$ and $\tilde m_0$ can be viewed 
as O($\alpha_{em}$) quantities. Expanding the r.h.s. of eqs.~(\ref{VP})--(\ref{SP}) 
in $\alpha_{em}$ to leading order we find
{\small{\beqn
\hspace{-.8cm}&&\langle V^1_0(x)P^2(0)\rangle^{latt}_{M_0} = ({\alpha_{em} \delta_{em}a^{-1}\!-\!\delta m_0)} \!\int\!\! d^4z \langle V^1_0(x)P^2(0) \bar\psi i\gamma_5\tau_3\psi(z)\rangle|^{L_{4}^{{QCD}}} +
  \,{\mbox{O}}({a})
\label{VP1}\\
\hspace{-.8cm}&&\langle S^1(x)P^1(0)\rangle^{latt}_{M_0} = {(\alpha_{em} \tilde\delta_{em}a^{-1}\!-\!\tilde m_0)} \!\int\!\! d^4z \langle S^1(x)P^1(0) \bar\psi i\gamma_5\psi(z)\rangle|^{L_{4}^{{QCD}}} +
 \,{\mbox{O}}({a}) \, ,
\label{SP1}
\eeqn}}  
\noindent
with all O($\alpha_{em}^2$) contributions in the r.h.s neglected. Hence
the contributions from the correlators in the r.h.s.\ of eqs.~(\ref{VP})
and~(\ref{SP}) are evaluated in the continuum $L^{QCD}_4$ theory 
(those in the first line, which are  O($\alpha_{em}$)) or 
ignored (those in the second line, which are O($\alpha_{em}^2$)). 
On the other hand the lattice correlators in the l.h.s.\ of 
eqs.~(\ref{VP1})-(\ref{SP1}) must be computed to O($\alpha_{em}$) in LQ(C+E)D.
We thus see that enforcing the conditions~(\ref{CC1}) allows 
to set $a\delta m_0 $ and $a\tilde m_0$ equal to $\alpha_{em} \delta_{em}$ and
$\alpha_{em} \tilde\delta_{em}$, respectively, thereby determining $M_{cr}$
(eqs.~(\ref{MCRF}) and (\ref{MCR2})) at the desired order. 

\section{LIB effects in Q(C+E)D with unquenched $u$, $d$, $s$ and $c$: 
a simulation strategy}
\label{sec:SISTRA}

Due to e.m.\ gauge invariance the lattice action for maximally twisted quark 
pairs $\psi^{\ell} = (u,d)$ and $\psi^{h} = (c,s)$ should in principle 
involve only identity and $\tau^3$ Pauli matrices in flavour space, i.e.\
\beq
S^{\ell h}_{33} = a^4 \sum_x\, \Big\{ \bar\psi^\ell(x)\Big{[}
\gamma\cdot{\widetilde\nabla}-i\gamma_5{\tau_3 W_{\rm{cr}} }
+\mu_\ell+{\tau_3} \epsilon_\ell\Big{]}\psi^\ell(x) \, + \,
\bar\psi^h(x)\Big{[}
\gamma\cdot{\widetilde\nabla}-i\gamma_5{\tau_3 W_{\rm{cr}} }
+\mu_h+{\tau_3} \epsilon_h \Big{]}\psi^h(x) \Big\} \, .
\label{SfermND} \eeq
As remarked above, the occurrence of a $i\gamma_5 \tau^3$ structure 
in the twisted critical Wilson term leads to a complex fermionic determinant, 
making problematic direct Monte Carlo simulations of this lattice formulation 
of Q(C+E)D~\footnote{In the absence of e.m.\ interactions one can use a
lattice action (which we might call 
$S^{\ell h}_{13}$) having instead a $i\gamma_5 \tau^1$ matrix in 
the twisted critical Wilson term: in this case the Dirac operator is 
$\gamma_5\tau^3$-Hermitean and has a real 
determinant~\cite{Frezzotti:2003xj}.}. The use of the RM123 method for
LIB corrections to physical observables substantially alleviates this
problem because the relevant info is extracted from renormalized 
correlation functions with insertions of suitable operators (e.m.\ currents,
isospin breaking quark densities)   
to be evaluated in the isosymmetric theory~\cite{deDivitiis:2013xla}, v.i.z. 
\beq
\!\langle \hat{\cal O} \rangle|_{\hat g_i}^{latt} = 
\langle \hat{\cal O} \rangle|_{\hat g_i^0}^{latt} \!\! + 
\hat e^2 \frac{\partial \langle \hat{\cal O} \rangle}{\partial \hat e^2} \Big|_{\hat g_i^0}^{latt} \!\! + 
\Big[ \hat g_s^2 - \Big(\frac{Z_{g_s}}{Z^0_{g_s}} \hat g_s^0 \Big)^2 \Big] \frac{\partial \langle \hat{\cal O} \rangle}{\partial \hat g_s^2}\Big|_{\hat g_i^0}^{latt} +
\Big[ \hat \mu_f - \Big( \frac{Z_{\mu_f}}{Z^0_{\mu_f}} \hat \mu_f^0 \Big) \Big]
\frac{\partial \langle \hat{\cal O} \rangle}{\partial \hat \mu_f}\Big|_{\hat g_i^0}^{latt} 
%
%
\, , \label{RM123-EXP}
\eeq
where $\hat g_i^0$ ($\hat g_i$) denote the renormalized couplings of the
isosymmetric (full) theory ($g_i = \{e, g_s, \mu_f \}$, with $\mu_f$ being 
the $f$-quark mass (at maximal twist), and $\hat{\cal O}$ are 
renormalized multilocal operators. The derivatives with respect to renormalized
couplings give rise to space-time integrated insertions of the corresponding
renormalized operators in the correlation functions of the isosymmetric theory. 
The maximally twisted mass regularization of the latter admits indeed a 
real positive fermionic determinant in the ($u,d$)-sector (where now
$\epsilon_\ell =0$) and, provided a suitable mixed action approach 
(see below) is employed, also in ($c,s$)-sector, where 
the large mass splitting term $\propto \epsilon_h$ occurs.

The evaluation of the renormalized correlation functions of interest should be
in principle carried out in the theory with action $S_{33}^{\ell h}$
at $e = \epsilon_\ell =0$. But having still $\epsilon_h \neq 0$ the simulation is
made problematic by the complex ($c$-$s$ sector) fermionic determinant.
However, up to negligible O($a^2$) lattice artifacts, precisely the same set
of renormalized correlation functions 
can also be evaluated by using the isosymmetric lattice action 
({\em mixed action approach})
\beqn
& S^{mix} \! = & S_{YM}(U,E)\Big|_{e=0} \!\! + 
S_{33}^{\ell h}(\mu_\ell, 0, \mu_h, \epsilon_h)\Big|_{e=0} \!\! +  
a^4 \sum_x \Phi_{gh}^{\dagger} \Big{[}
\gamma\cdot{\widetilde\nabla}-i\gamma_5{\tau_3 W_{\rm{cr}} }
+\mu_h+{\tau_3} \epsilon_h \Big{]} \Phi_{gh}\Big|_{e=0}  + \nonumber \\
& & +\;  a^4 \sum_x \bar\psi^h_{sea}(x)\Big{[}
\gamma\cdot{\widetilde\nabla}-i\gamma_5{\tau_1 W_{\rm{cr}} }
+\mu_h+{\tau_3} \epsilon_h^{sea} \Big{]}\psi^h_{sea}(x)\Big|_{e=0}
\label{SLATMIX}
\eeqn  

\vspace{-0.1cm}
\noindent where $S_{YM}(U,E)$ denotes the gluon and photon lattice action and
$S_{33}^{\ell h}$ is given in eq.~(\ref{SfermND}). The virtual ``sea'' 
effects of the quark fields $\psi_h$, $\bar\psi_h$ are canceled by those of
the complex ghost field $\Phi_{h}$ but reintroduced 
through the fields $\psi_h^{sea}$, $\bar\psi_h^{sea}$. 
Of course the mixed action theory $S^{mix}$ must be simulated at 
(fixed as $a \to 0$) renormalized parameters ($\hat g_i^0$) matching those of the
theory $[S_{YM}(U,E) + S_{33}^{\ell h}(\mu_\ell, 0, \mu_h, \epsilon_h)]|_{e=0}$ 
one should have studied in principle, and in particular with equal values for the 
valence and sea renormalized masses of each quark flavour. 
In the isosymmetric theory this implies working with equal values 
of the valence and sea bare mass parameters, except for
$\epsilon_h = Z_{P^0} Z_S^{-1} \epsilon_h^{sea}$. Only e.m.\ corrections will
give rise to the more complicated pattern of eq.~(\ref{MASSREN2}). 

The proof of the statements above is straightforward~\cite{PREP} and can be given 
along the lines of ref.~\cite{Frezzotti:2004wz}.
We note that the form of the valence $c$-$s$ sector in the action~(\ref{SLATMIX}) 
is such that within the RM123 approach for LIB corrections to hadronic observables
one has to evaluate precisely the same
set of Wick contractions that should be computed if the action $S_{33}^{\ell h}$
were adopted, but employing gauge ensembles that are generated with  
no ``complex phase problems''. This is so because $\epsilon_\ell =0$ and the $c$-$s$ 
sea quark effects stem from the action term in second line of eq.~(\ref{SLATMIX}). 

\vspace{0.15cm}
\noindent
{\bf Acknowledgements} We thank V.~Lubicz, G.~Martinelli and S.~Simula for
stimulating discussions. R.F.~and~N.T. acknowledge support through
the grant ``Uncovering Excellence 2014 - LIBETOV''.

%

\end{document}